\def\Para#1{}

\documentstyle[aps,prl,twocolumn,epsfig,floats]{revtex}

\begin{document}
\bibliographystyle{prsty}

\draft 

\title{Low-Temperature Saturation of the Dephasing Time  \\ 
and Effects of Microwave Radiation
 on Open Quantum Dots } 

\author{A. G. Huibers, J. A. Folk, S. R. Patel, and C. M. Marcus}
\address{Department of Physics, Stanford University, Stanford, California 94305}

\author{C. I. Duru\"oz and J. S. Harris, Jr.}
\address{Department of Electrical Engineering, Stanford University, Stanford, California
94305}

\date{\today}

\maketitle

\begin{abstract} The dephasing time $\tau_\varphi$ of
electrons
 in open semiconductor quantum dots, measured using ballistic weak localization, is found to
saturate below
$\sim 100 mK$, roughly twice the electron base temperature, independent of dot size. Microwave radiation
deliberately coupled to the dots affects quantum interference indistinguishably from elevated temperature,
suggesting that direct dephasing due to radiation is not the cause of the
observed saturation. Coulomb blockade measurements show that the applied microwaves
 create sufficient source-drain voltages to account for dephasing due to Joule heating.
\end{abstract}


Phase coherent electronics have attracted considerable interest both because of their
potential use as the basis of solid-state quantum information processing \cite{DiVincenzo},
and because of significant discrepancies between theory and experiment on the subject of
low-temperature dephasing \cite{Mohanty97,Aleiner98x}.  Dephasing, or quantum decoherence, results from interactions
between a quantum system and its environment. In the context of mesoscopic physics 
\cite{Imry97}, the quantum system is an electron or quasiparticle in a conductor and the
environment includes phonons, radiation, magnetic impurities, and other electrons.

The theory of dephasing in low-dimensional conductors, as well as methods for
extracting the time scale on which dephasing occurs, $\tau_\varphi$, from
transport measurements are by now well established 
\cite{Altshuler85}. Measurements of $\tau_\varphi$ in a variety of one- and
two-dimensional (1D, 2D) disordered metals and semiconductors
\cite{Mohanty97,Liu86,Kurdak92,Gershenson98}, clean 2D
semiconductors \cite{Yacoby91}, and zero-dimensional (0D)
quantum dots \cite{BirdClarke,Huibers98} have been reported, with $\tau_\varphi$
typically in the range of picoseconds to tens of nanoseconds, depending on temperature. Interestingly,
essentially all experiments show some saturation of $\tau_\varphi$ at low temperatures, the origin of which
remains unresolved. This issue is quite important since an intrinsic saturation of
$\tau_\varphi$ at low temperatures would signal a breakdown of Fermi-liquid behavior.

\begin{figure}[bth]
\epsfxsize=3 in \epsfbox{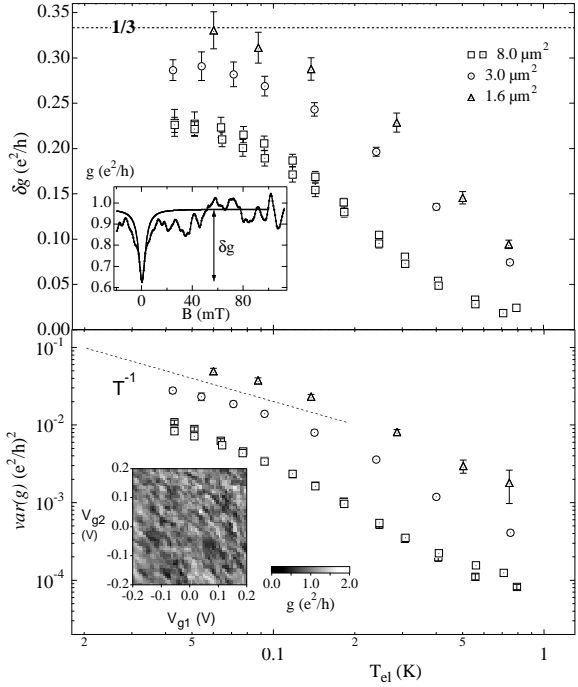}
\vspace{2mm}
\caption{ (a) Weak localization amplitude $\delta g$ as a function of electron temperature $T_{el}$ in dots with
single mode leads and area $A = 1.6\;\mu m^2$ (triangles), $3.0\;
\mu m^2$ (circles), $8.0\; \mu m^2$ (squares). Inset: Average conductance
$\langle g(B) \rangle$ as a function of magnetic field $B$, from 20 traces for a $1.0\; \mu m^2$ device, showing
weak localization at $B=0$, with Lorentzian fit (solid). (b) Variance of conductance fluctuations, $var(g)$, as a
function of
$T_{el}$ for the same devices, measured with broken time-reversal symmetry, $B >
\Phi_0/A$. $T_{el}$ is measured from CB peak width in the  $8.0\;\mu m^2$ device.  Both $\delta g$ and $var(g)$ are 
based on ensembles of $\sim 400$ independent values of $g$ drawn from shape distortion landscapes. Lower inset
shows a shape-distortion landscape for the $1.6\; \mu m^2$ device.}
\label{fig1}
\end{figure}

\begin{figure}[bth]
\epsfxsize=3 in \epsfbox{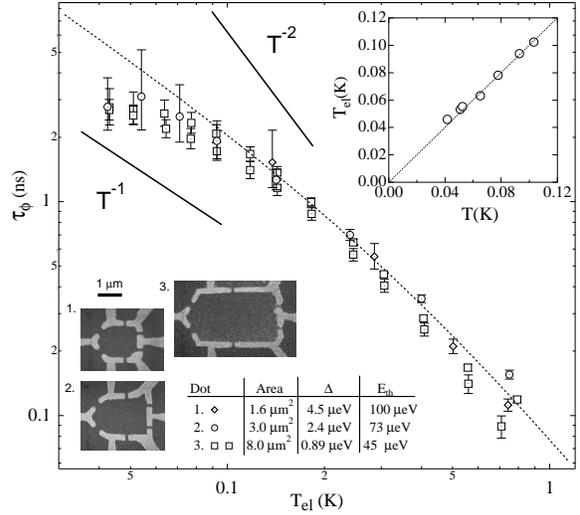}
\vspace{2mm}
\caption{ Phase coherence time $\tau_\varphi$ as a function of $T_{el}$, 
determined from $\delta g$, as described in text, along with empirical relation $\tau_\varphi[ns] = (
4.\;T[K]+ 9.\;T[K]^2)^{-1}$.  Lines show $T^{-1}$ and $T^{-2}$ dependencies. Upper inset: Electron temperature
from CB peak width as a function of mixing chamber temperature, $T$, in the $3.0 \mu m^2$ device. Lower inset:
micrographs and device parameters, with ballistic Thouless energy defined $E_{th} = \hbar v_F/L$, where $L$ is the
width of the device.}
\label{fig2}
\end{figure}

Quantum dots---small  islands of charge connected to electronic reservoirs via conducting leads---are
particularly useful for studying coherence effects because quantum corrections to the
conductance of the dot are large, comparable to the average conductance itself.
Another useful feature is that the same device that is used to
measure 
$\tau_\varphi$ can also be used to measure the electron temperature, $T_{el}$, either via Coulomb blockade (CB)
peak widths
\cite{Kouwenhoven97} or from the variance of conductance
fluctuations, $var(g)$, the latter taking advantage of results from random matrix
theory (RMT) \cite{RMT} and nonlinear sigma model calculations \cite
{Efetov95} that assume the dot to be either disordered or chaotic.

This Letter contains two main results: First, we present evidence of a low-temperature saturation of
$\tau_\varphi(T_{el})$ below $T_{el}\sim 100 mK$ in clean quantum dots with single channel leads, based on
measurements of ballistic weak localization. We further find that
$\tau_\varphi(T_{el})$ is independent of dot
size and shape throughout the measured temperature range, consistent with previous measurements \cite{Huibers98}. 
Low-temperature saturation of $\tau_\varphi$ in quantum dots has been reported previously
\cite{BirdClarke} based on less precise measurement methods or methods requiring sizable magnetic
fields. Second, we investigate the possibility that the cause of saturation of $\tau_\varphi$ is unintentional
irradiation, which can cause direct dephasing at amplitudes much too weak to induce heating
\cite{Altshuler82,Aleiner98x}. Direct dephasing by microwave radiation, mixed with varying degrees of Joule heating,
has been observed in experiments on disordered wires \cite{Liu90}, films
\cite{Liu91,Wang87,Bykov88} and MOS structures
\cite{Vitkalov88}. Neither theory nor previous experiment has addressed the
case of quantum dots to our knowledge. Our investigation consists of applying microwave radiation to the sample
over a range of amplitudes at frequencies from well above to well below
$2\pi/\tau_\varphi$. We find that the radiation indeed reduces $\tau_\varphi$, but that the reduction
appears well accounted for by electron heating rather than direct dephasing. From this observation we conclude that
the saturation of
$\tau_\varphi(T_{el})$ in the absence of applied radiation cannot be easily explained by  unintentional
microwave irradiation. Similar conclusions were recently reached  by Webb and coworkers
\cite{Webb_99} based on similar microwave irradiation experiments on metal wires.

\begin{figure}[bth]
\epsfxsize=3 in \epsfbox{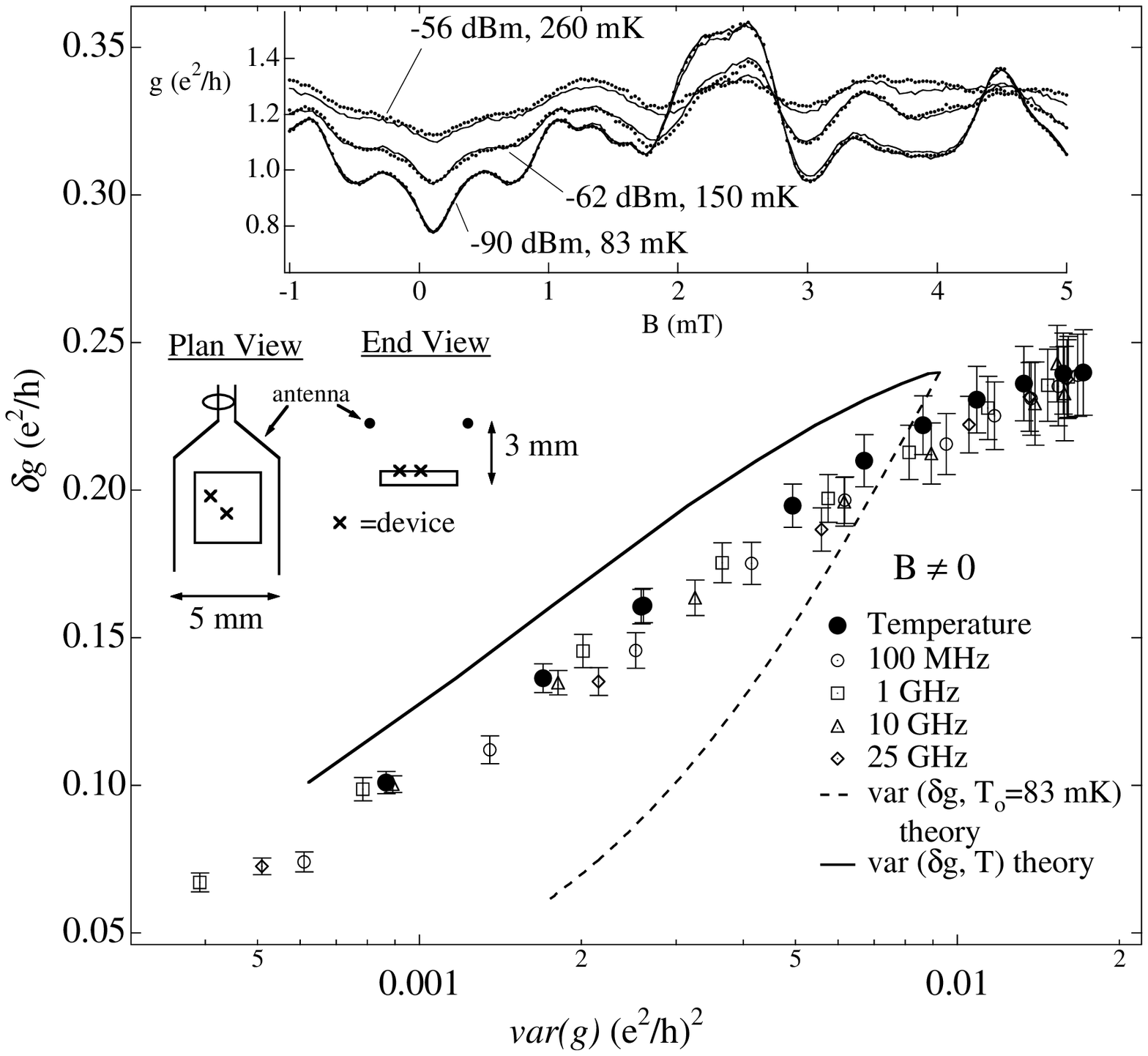}
\vspace{2mm}
\caption{Weak localization $\delta g$ versus conductance fluctuations $var(g)$ for the
$3.0\;\mu m^2$ dot parameterized by both $T_{el}$ (solid squares) and microwave power at base temperature
$T_{el}$=83 mK, for frequencies ranging from 100 MHz to 25 GHz. Curves show RMT results for
dephasing without heating at 83mK (dashed curve) and dephasing due to heating (solid curve)
using measured values of
$\delta g$ as input to find $T_{el}$ and hence
$var(g)$. Inset: single magnetoconductance traces at
different microwave power levels at 700 MHz for base temperature (dotted curves) along with traces taken for
increasing electron temperature (solid curves). In all cases, effects of heating and irradiation are
indistinguishable.}
\label{fig3}
\end{figure}

Measurements are reported for five devices, each formed using lithographically patterned gates 90 nm
above a two dimensional electron gas (2DEG) in a delta-doped $GaAs/Al_{0.3}Ga_{0.7}As$
heterostructure (Fig.\ 2 insets). A sheet density $ \sim 2 \times 10^{11}\;cm^{-2}$
and mobility
$1.4
\times 10^5\;cm^2/Vs$ give a mean free path of
$\sim 1.5\;\mu m$, comparable to the device size (see Fig. 2, table). Measurements were carried out in a dilution
refrigerator after slowly cooling the samples with a bias  of $+ 0.4$ V on all gates to reduce switching noise. The
base $T_{el}$ was 45 mK in the saturation experiments (i.e. before adding external microwave apparatus) and
83 mK in the microwave irradiation experiments, which required the removal of a low-temperature shield. The base
mixing chamber temperature was 28 mK. $T_{el}$ was determined from CB peak widths measured at several
temperatures \cite{Kouwenhoven97}. All conductances were measured using analog lock-in amplifiers (PAR 124) with
less than $2
\mu\rm V$ $(< kT/e)$ across the dot.  Statistics of conductance data were gathered by sampling an
ensemble of shape distortions controlled by two gates $V_{g1}$ and $V_{g2}$ (Fig. 1 inset) while simultaneously
trimming the point contact gate voltages to maintain precisely one mode in each lead \cite{Huibers98}.

Dephasing time was extracted from the change in ensemble-averaged conductance, $\delta g = (\langle g\rangle_{B\neq
0} - \langle g\rangle_{B=0})$, upon breaking time reversal symmetry with a small magnetic field 
\cite{Huibers98}. Noninteracting RMT \cite{RMT} yields a simple approximate expression for
$\delta g$ that depends only on dephasing and the number of channels in the leads,
\begin{equation}
 \delta{\rm g} \sim 1/(2N + 1 + \gamma_\varphi)\;e^2/h,
\end{equation} where $\gamma_\varphi =2\pi
\hbar / (\tau_\varphi \Delta)$ is the dimensionless dephasing rate (i.e.\ the number of dephasing channels),
$\Delta  ={2\pi \hbar ^2} / {m^*A}$ is the mean level  spacing, and $A$ is the dot area. Equation (1) allows
$\tau_\varphi$ to be extracted from
$\delta g$ knowing only the area of the dot and the conductance of the leads. Figure 1(a) shows that the
zero-dephasing limit for one-mode leads ($N=1$),
$\delta g = 1/3\;e^2/h$, is approached in the smaller devices, indicating that $\gamma_\varphi < 1$ in these
devices. Because
$\gamma_\varphi$ depends on $A$, the same values of
$\tau_\varphi$ in the larger devices yield large values of $\gamma_\varphi$, and
correspondingly smaller $\delta g$, according to Eq. (1). 

Conductance fluctuations are also reduced by dephasing, but, unlike
$\delta g$, are further reduced by thermal averaging
\cite{RMT,Huibers98b}. For broken time-reversal symmetry and $kT > \Delta$, the RMT result including
thermal averaging can be well approximated  by
$
var(g) \sim (\Delta/ {6 k T})  f(\gamma_\varphi),
$
where $f(\gamma_\varphi) = 2 (2 + \gamma_\varphi)^{-1} (\sqrt {3} + \gamma_\varphi)^{-2}$,
as discussed in Ref.\ \cite{Huibers98b}. Notice that for $\gamma_\varphi < \sim 1$,  $var(g) \propto T^{-1}$, so
that
$var(g)$ serves as a low-temperature electron thermometer. Values for $T_{el}$ obtained from $var(g)$ are
consistent with values obtained from CB peak widths over the full temperature range of the experiment.

Figure 2 shows the phase coherence time $\tau_\varphi$, extracted from $\delta g$ using Eq.\ (1), as a
function of
$T_{el}$ measured from CB peak widths.  Above $T_{el}\sim 100$ mK, a dephasing rate of the form
$\tau_\varphi^{-1} \sim
A\;T_{el} + B\;T_{el}^2$ is observed, consistent with previous experimental results \cite{Huibers98}, but not
expected theoretically \cite{SIA}. Below 100 mK,
$\tau_\varphi$ appears to saturate.


Electron-electron interaction effects not accounted for within RMT are expected to
enhance
$\delta g$ by a factor of 
$\sim (1 + 0.24 \Delta/kT)$ at $N=1$, as well as enhance $var(g)$ by a factor
$\sim (1+
 2\Delta /kT)$ for  $N \gg 1$, according to recent theory \cite{Brouwer98}. Variance
results for
$N=1$ are not known, nor are the effects of dephasing on these enhancements \cite{Brouwer98}. Nonetheless,
these results suggest that interaction effects are small, particularly in the larger
dots. For instance, for the 8
$\mu m^2$ device interactions enhance
$\delta g$ by $\sim$ 5\% at 45mK and less at higher temperatures. More importantly, since the enhancement of
$\delta g$ increases as temperature decreases, interaction affects alone cannot account for the
observed saturation.

To investigate possible causes of the saturation of $\tau_\varphi(T_{el})$, a set of quantum dots were 
deliberately  irradiated with microwaves at frequencies ranging from well above to well below $\tau_\varphi^{-1}$.
The aim was to intensify any direct dephasing without heating so it could be unambiguously observed. Microwaves were
coupled into the dilution refrigerator via a 1.5 mm coaxial waveguide attached to an open biaxial antenna segment
of width 5 mm, positioned 3 mm above the sample (Fig.\ 3, inset). Note that these dimensions are much smaller than
the radiation wavelength ($\lambda=30$ cm$/f$[GHz]).

Figure 3 shows that the parametric dependence of $\delta g$ (sensitive to dephasing only) versus $var(g)$ (sensitive
to temperature as well as dephasing) evolves the same when either the temperature or microwave
power is increased. This would not be the case if the microwaves caused dephasing without heating: The dashed
curve indicates the parametric dependence for a variable dephasing rate with $T_{el}$ fixed at the base electron
temperature, 83 mK, corresponding to dephasing without heating. On the other hand, the solid curve allows both
temperature and dephasing to vary, using $\delta g$ to determine
$T_{el}$ (from Fig.\ 1). The qualitative agreement with this solid curve as well as the overall
indistinguishability of microwave radiation from heating indicate that the only observable effect of the
microwaves on electrons in open dots is heating. This conclusion is further supported by the
individual magnetoconductance traces at elevated temperatures (Fig. 3, inset). Feature by
feature, these traces are indistinguishable from traces taken at base temperature with an
appropriate microwave power applied.

Having established that the applied microwaves appear to cause dephasing through heating, we
 can further infer the likely heating  mechanism by examining the dependence of the effective electron temperature
in the dot on microwave power, as well as rectification in the CB regime. Together, these data suggest that the
applied microwaves induce microvolt-scale ac source-drain voltages sufficient to cause Joule heating within the
dot. We have previously found that the temperature $T_{dot}$ inside a single-mode ($N=1$) quantum dot in the
presence of a finite dc source-drain bias, $V_{sd}^{(dc)}$, is well characterized by balancing Joule heating
and out-diffusion of hot electrons, giving 
$T_{dot} = (T_{res}^2 +
\alpha (3\pi^2 e^2/2 k^2) V_{sd}^2)^{1/2}$, 
where $T_{res}$ is the reservoir electron temperature and $\alpha
\sim 1/2$ is the fraction of heat that flows into the dot
\cite{Switkes98}. 

\begin{figure}[bth]
\epsfxsize=\columnwidth \epsfbox{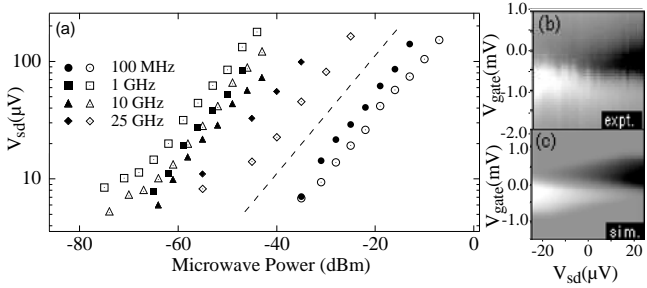}
\vspace{1mm}
\caption{(a) Estimated ac source-drain voltage $V_{sd}$ across the dot as a function of the microwave power,
$P$. Dot temperature used to infer $V_{sd}$ (see text) is from either CB peak widths (filled points)
or $\delta g$ (open points) from the $3.0 \mu m^2$ dot. Dashed line shows the expected slope
$V_{sd} \propto P^{1/2}$, for the case of Joule heating balanced by out-diffusion of hot electrons. (b,c) Grayscale
plots of (b) experimental (c) simulated dc current
in the vicinity of a CB ``diamond", as a function of gate voltage and a dc bias across the dot. Skewed branches
of the diamond reflect ac biases induced by the microwaves.  (b) Experiment: base temperature data for a $1.0
\mu m^2$ dot, microwave power at -37 dBm at 25 GHz. (c) Simulation: ac source-drain fluctuations with rms amplitude
of $30 \mu V$.}
\label{fig4}
\end{figure}

Figure 4(a) shows the effective ac source-drain voltage $V_{sd}^{(ac)}$, extracted from $T_{dot}$ and
$T_{res}$, as a function of microwave power, $P$. The value for $T_{dot}$  was found by two methods, first by
matching 
$\delta g$ and $var(g)$ data (Fig.\ 3) with those for 
increasing temperature, and second from CB peak widths.  Figure 4(a) shows the expected dependence,
$V_{sd}^{(ac)} \propto P^{1/2}$ from Joule heating plus out-diffusion, but does not
rule out other heating mechanisms such as direct absorption of radiation by electrons in the
dot. Variablility in the coupling of the radiation to the 2DEG allows an overall horizontal
shift in the data. At some frequencies (e.\ g.\ $100$ MHz and $10$ GHz) the two methods track
closely, whereas at other frequencies the two methods give curves with an overall shift for
unknown reasons.

Direct evidence that heating due to microwaves results from an ac source-drain voltage can be
seen in the  CB ``diamonds", shown in Fig.\ 4(b) as grayscale plots of current as a function of
$V_{sd}^{(dc)}$ and gate voltage $V_g$.  Notice that a nonzero current flows at $V_{sd}^{(dc)} = 0$ (first
positive, then negative, as
$V_g$ is swept), i.e. the diamonds do not meet at a point but are skewed. This skewing {\it cannot} result from
temperature alone, and provides a direct measure of $V_{sd}^{(ac)}$ \cite{Huibers99}.
Simulations that include a dc + ac source-drain voltage and typically
asymmetric CB diamonds
\cite{Kouwenhoven97} show a similar skewing, as seen in Fig. 4(c).

In summary, we have measured electron dephasing in quantum dots ranging in size over a factor of 5, and observe a
$A\;T_{el} + B\;T_{el}^2$ dependence of the dephasing rate with an unexplained saturation below $\sim$ 100 mK.
Microwave radiation applied to the dots appears to cause dephasing only by heating, presumably through modulation
of the source-drain voltage.  Therefore, microwave coupling of the type investigated here does not appear to
explain the observed saturation. It would interesting in future experiments to compare broadband versus
monochromatic radiation \cite{Stern90}, radiation specifically coupled to the confining gates, and far-field or
cavity radiation where $E$ and $B$ are of equal amplitude.

We thank I. Aleiner, B. Altshuler, M. Jarawala, U. Sivan, A. Stern, and F. Zhou for useful discussions. We 
gratefully acknowledge support from the Army Research Office under Grant DAAH04-95-1-0331 the NSF-PECASE
programs (Marcus Group), Hertz Foundation (AGH), and JSEP under Grant DAAH04-94-G-0058 (Harris Group).


\begin{thebibliography}{10}
\bibitem{DiVincenzo}  D. Loss, and D. P. DiVincenzo, Phys. Rev. A {\bf 57}, 1050 (1998); G. Burkard, D. Loss, and D.
P. DiVincenzo, Phys. Rev. B {\bf 59}, 2070 (1998).

\bibitem{Mohanty97} P. Mohanty, E. M. Q. Jariwala, and R. A. Webb, Phys. Rev. Lett. {\bf 78}
3366 (1997); P. Mohanty and R. A. Webb, Phys. Rev. B. {\bf 55} R13452 (1997).

\bibitem{Aleiner98x}I.L. Aleiner, B.L. Altshuler, and M.E. Gershenson, {\it Interaction effects and phase
relaxation in disordered systems}, cond-mat/9808053 (1998).

\bibitem{Imry97}For an introduction to the subject, see: Y.~Imry {\it An Introduction to Mesoscopic Physics}
(Oxford Univ. Press, Oxford, 1998).

\bibitem{Altshuler85} B. L. Altshuler and A. G. Aronov, in {\it Electron-Electron Interaction in
Disordered Systems}, edited by A. L. Efros and M. Pollak (Elsevier, Amsterdam, 1985).

\bibitem{Liu86} J. J. Liu and N. Giordano, Phys. Rev. B {\bf 33}, 1519
(1986); K. K. Choi, D. C. Tsui, and K. Alavi, Phys. Rev. B {\bf 36},
7551 (1987);

\bibitem{Kurdak92} C. Kurdak {\it et~al.}, Phys. Rev. B
{\bf 46}, 6846 (1992); P. M. Echternach, M. E. Gershenson, and H. M.
Bozler,  Phys. Rev. B {\bf 48}, 11516 (1993).

\bibitem{Gershenson98} J. Katine {\it et~al.}, Phys. Rev. B {\bf 57},
1698 (1993); M.~E. Gershenson {\it et~al.}, Phys.  Rev.  Lett.  {\bf
81}, 1066 (1998).

\bibitem{Yacoby91} A. Yacoby {\it et~al.}, Phys. Rev. Lett. {\bf 66},
1938 (1991).

\bibitem{BirdClarke} J. P. Bird {\it et~al.}, Phys. Rev. B {\bf 51},
18037 (1995); R. M. Clarke {\it et~al.}, Phys. Rev. B {\bf 52}, 2656
(1995).

\bibitem{Huibers98} A.~G. Huibers {\it et~al.}, Phys.  Rev.  Lett.  {\bf
81}, 200 (1998).

\bibitem{Kouwenhoven97} L.P. Kouwenhoven {\it et~al.}, in {\it
Mesoscopic Electron Transport}, edited by L.L. Sohn, L.P.  Kouwenhoven
and G.  Sch\"on (Kluwer, Dordrecht, 1997).

\bibitem{RMT} H.~U. Baranger and P.~A. Mello, Phys.  Rev.  B {\bf 51}
1995; P.~W. Brouwer and C.~W.~J. Beenakker, Phys. Rev.  B {\bf 55}, 4695
(1997).

\bibitem{Efetov95} K.B. Efetov, Phys.  Rev.  Lett.{\bf 74}, 2299 (1995).

\bibitem{Altshuler82} B. L. Altshuler {\it et~al.}, J. Phys. C {\bf
15}, 7367 (1982).

\bibitem{Liu90} J. J. Liu and N. Giordano, Phys. Rev. B {\bf 41}, 9728
(1990).

\bibitem{Liu91} J. J. Liu and N. Giordano, Phys. Rev. B {\bf 43}, 1385
(1991).

\bibitem{Wang87} S. Wang and P.~E. Lindelof, Phys.  Rev.  Lett.  {\bf
59}, 1156 (1987); S. Wang and P.~E. Lindelof, J. Low Temp. Phys.
{\bf71}, 403 (1988).

\bibitem{Bykov88} A. A. Bykov, G. M. Gusev and Z. D. Kvon, J. Phys. C
{\bf21}, L585 (1988).

\bibitem{Vitkalov88} S.~A. Vitkalov {\it et~al.}, Sov. Phys. JETP {\bf
67}, 1080 (1988).

\bibitem{Webb_99} P. Mohanty, E. M. Q. Jariwala, and R. A. Webb,
preprint.


\bibitem{Huibers98b} A.~G. Huibers {\it et~al.}, Phys.  Rev.  Lett. 
{\bf 81}, 1917 (1998).

\bibitem{SIA} U. Sivan, Y. Imry, and A. G. Aronov, Europhys.\ Lett.\
{\bf 28}, 115 (1994).

\bibitem{Brouwer98} P.~W. Brouwer and I.~L.~Aleiner, Phys.  Rev.  Lett. 
{\bf 82}, 390 (1999).

\bibitem{Switkes98} M. Switkes {\it et~al.}, Appl. Phys. Lett.  {\bf
72}, 471 (1998).

\bibitem{Huibers99} A.~G. Huibers, Ph.D. Thesis, Stanford University,
1999.

\bibitem{Stern90} A. Stern, A. Aharonov, and Y. Imry, Phys. Rev. A {\bf 41},
3436 (1990).

\end{thebibliography}
\end{document}